%Paper: hep-ph/9211289
%From: yuan@msupa.pa.msu.edu
%Date: Fri, 20 Nov 1992 16:10:48 EST
%Date (revised): Mon, 1 Mar 1993 18:19:16 -0500 (EST)

\input phyzzx
\overfullrule=0pt
%========================DEFINITIONS===================================
\def\ETslash{\not{\hbox{\kern-4pt $E_T$}}}
\def\mynot#1{\not{}{\mskip-3.5mu}#1 }

\def\ra{\rightarrow}

\def\tbW {$t$-$b$-$W$ }

\def\width{ \Gamma( t \ra b W^+) }
\def\journal#1&#2{#1{\bf #2}}

\nopubblock
\line{\hfil MSUTH 92/05 }
%\line{\hfil November, 1992}
\line{\hfil February, 1993}
\line{\hfil revised}

\titlepage
\title{ Studying the Top Quark via the $W$--Gluon Fusion Process}
\author{ Douglas O. Carlson \ \ and  \ \ \ C.--P. Yuan}
\address{
Department of Physics and Astronomy \break
Michigan State University  \break
East Lansing, MI 48824}

\singlespace
\abstract{
We show that studying the single top quark production via the
$W$--gluon fusion process can provide unique information on:
({\it i}) the measurement of the decay width $\width$;
({\it ii}) probing the symmetry breaking
mechanism by measuring the form factor of~\tbW;
({\it iii}) testing the Effective--$W$ Approximation prior to supercolliders;
({\it iv}) testing CP violation by observing
different production rates of $\bar p p \ra t X $ and $\bar p p \ra \bar t X$;
({\it v}) testing CP violation from the almost hundred percent
longitudinally polarized top quark decay; in addition to the measurement of the
top quark mass.
In contrast to the recent claim by R. K.~Ellis and S.~Parke, we show that this
process is extremely useful at the Tevatron.
}

\endpage
%=========================Body of paper=======================
\normalspace
\chapter{ Introduction }

The top quark has been found to be heavier than 45 GeV from
SLAC and LEP experiments
and 91 GeV from CDF data.
The first limit is model independent while the second limit is for the
Standard Model (SM) top quark. Since the top quark is heavy, of the
same order of magnitude as the $W$--boson mass, any physical observable
related to the top quark may be sensitive to new physics.
The top quarks will therefore allow many new tests of the SM and new probes of
physics at the 100 GeV scale.

\REF\wgone{C.--P. Yuan, Phys. Rev. {\bf D41} (1990) 42.}
\REF\rnew{S. Dawson, Nucl. Phys. {\bf B249}, 42 (1985);
S. Dawson and S. S. D.Willenbrock, {\bf 284}, 449 (1987);
S. S. D. Willenbrock and D. A. Dicus, Phys. Rev. D {\bf 34}, 155 (1986);
F. Anselmo, B. van Eijk and G. Bordes, Phys. Rev. D {\bf 45}, 2312 (1992);
T. Moers, R. Priem, D. Rein and H. Reithler in Proceedings of Large
Hadron Collider Workshop, preprint CERN 90-10, 1990.}

In the past, one of us proposed a method, via the $W$--gluon fusion
process $q g \, (W^+g) \ra t \bar{b} X$, to detect a heavy top quark at hadron
colliders.\refmark{\wgone,\rnew}
We showed that an almost perfect efficiency for the ``kinematic $b$ tagging''
can be achieved
due to the characteristic features of the transverse momentum and
rapidity distributions of the spectator quark which emitted the virtual $W$.
For a 140 GeV SM top quark, the production rate ($\sim 4$ pb)
 for the $W$--gluon fusion process $Wg \ra (t \bar b + \bar t b)$ is
about a factor of 4 smaller than the usual QCD processes $q \bar q,
\, gg \ra t \bar t$ at the Tevatron. However, the $W$--gluon fusion process
becomes more important for a heavier top quark.

\REF\toppol{ G. L. Kane, G. A. Ladinsky and C.--P. Yuan,
Phys. Rev. {\bf D45} (1992) 124.}
\REF\doug{D.~Carlson and C.--P. Yuan, in preparation.}
\REF\effw{ For a review, see S. Cortese and R. Petronzi,
Phys. Lett. {\bf B276} (1992) 203.}

With a $100 \, {\rm pb}^{-1}$ luminosity at the
Tevatron, it is important to re--examine this process
to see whether it can teach us something which
is either impossible or difficult
to be learned from studying the usual QCD process.
Both CDF and D0 are upgrading detection efficiencies for forward jets,
 the coverage of leptons, and
the efficiency of tagging $b$ jets in their detectors.
Putting all of these together, we
show that studying this process can give a better measurement on
the decay width $\width$ of the top quark.
For instance, the top quark decay width $\width$ can be measured
better via this process at the Tevatron
than the usual QCD process at the supercolliders
by almost an order of magnitude for a 140 GeV top quark.\Ref\same{
The measurement of ~$\width$ is equivalent to the
measurement of~\tbW couplings.}
Therefore, it allows probing the symmetry breaking
mechanism by measuring the form factors of~\tbW
from its production rate.
This process also offers a chance to study the ``Effective--$W$
Approximation,''\refmark{\effw}
 which is essential in studying the strongly interacting
longitudinal $W$ system, prior to the time of the future
high--energy colliders SSC and LHC.
At the future hadron--hadron, electron--hadron, or $e^-e^+$
 colliders, one might be able to
use similar processes\refmark{\toppol} to measure the
Cabbibo--Kobayashi--Maskawa matrix element $|V_{ts}|$ due to the copious
production of the top quark.\refmark{\doug}

\REF\pxi{R.D.~Peccei and
X.~Zhang, \journal Nucl.~Phys. &{B337} (1990) 269.}

It is important to examine
different types of operators the~\tbW vertex might
have.\refmark{\toppol}
For instance, one should examine the form factors
which result from an effective lagrangian obtained by higher order
QCD and electroweak effects.
One may also examine the form factors to test the
plausibility of having the {\it nonuniversal} gauge couplings
of~\tbW due to some dynamical symmetry breaking scenario.\refmark{\pxi}

The most important consequence of a heavy top quark
is that to a good approximation it decays as a free quark since its
lifetime is short and it does not have time to
bind with light quarks before it decays. Thus we can use the polarization
properties of the top quark
as additional observables to test the SM and
to probe new physics.
 Furthermore, because the heavy top quark
has the weak two--body decay $t \ra b W^+$, it will
analyze its own polarization.

Top quarks will have longitudinal
polarization if weak effects are present in their production.
In the SM, the heavy top quark produced via the $W$--gluon
fusion process is left--handed polarized, and unpolarized from the usual
QCD process, at the Born level.
With the large production rate expected for top quarks at the SSC
and the LHC, it
will be possible to make a detailed study of the interactions of the top
quark including polarization effects.
If new interactions occur, they may manifest themselves in
an enhancement of the polarization effects in
the production of the top
quark via the $W$--gluon fusion process.\refmark{\toppol}
Furthermore, if CP is violated, the production rate of $t$ from
 $ \bar p p(W^+ g) \ra t \bar b X$
would be different from that of $\bar t$ from
$\bar p p(W^- g) \ra \bar t  b X$.
Therefore, one can detect large CP violation effects
by observing the difference in the production rates of $t$ and $\bar t$.

\REF\nucl{
J.D. Jackson, S.B. Treiman and H.W. Wyld, Jr., \journal
Phys.~Rev. & {106} (1957) 517;
R.B. Curtis and R.R. Lewis, \journal Phys.~Rev. & {107} (1957) 543.}
\REF\gunion{
B. Grzadkowski and J.F. Gunion, preprint UCD-92-7, 1992.}
\REF\parke{
R.K. Ellis and S. Parke, Phys. Rev. {\bf D46} (1992) 3785.}

In the $W$--gluon fusion process the top quark is
almost hundred percent longitudinally
polarized. This allows us to probe CP violation
in the decay process
$t \ra W^+ b \ra l^+ \nu_l b$.
The obvious observable for this purpose
is\refmark{\nucl}
the expectation value of the time--reversal quantity
$
\vec{\bf\sigma_t} \cdot (\widehat{\bf p}_b \times \widehat{\bf p}_{l})
$
where $\vec{\bf\sigma_t}$ is the polarization vector of $t$, and
$\widehat{\bf p}_b$ ($\widehat{\bf p}_l$) is the unit vector of the
$b$ ($l^+$) momentum in the rest frame of the top quark.
This was suggested in Ref.~\toppol\ and further studied in Ref.~\gunion.

In contrast to the recent claim by R.~K.~Ellis and S.~Parke,\refmark{\parke}
 we show in the next section that this
process is extremely useful at the Tevatron.
With kinematic cuts, there are about 20 $t$ or $\bar t$
(including the branching ratio $W \ra e, \, {\rm or} \, \mu$)
from this process produced
at the Tevatron with a 100 ${\rm pb}^{-1}$ luminosity and $\sqrt{S}=1.8$ TeV.
The major background is the $W+\, jets$ event which is of the same order as
the signal event with $b$--tagging using a vertex detector.  The $b$--quark
tagging efficiency is assumed to be 100\% with no misidentifications.

\chapter{ Brief Results }

In this letter, we will only briefly report our results and leave
the details to future publications.
For simplicity, we only discuss 140 GeV top quark production at the Tevatron.

\REF\rjohn{ J. C. Collins and C.--P. Yuan, in preparation.}
\REF\wuki{ J. C. Collins and Wu--Ki Tung, \journal Nucl.~Phys. &{B278}
(1986) 934;
F. Olness and Wu--Ki Tung, Nucl. Phys. {\bf B308} (1988) 813;
M.  Aivazis, F. Olness and Wu--Ki Tung, Phys. Rev. Lett.
{\bf 65} (1990) 2339.}
\REF\pdf{ J. Morfin and Wu--Ki Tung, \journal Z.~Phys. &{C52} (1991) 13.}

Since the single top quark produced from the $W$--gluon fusion process
involves a very important and not
yet well--developed technique of handling the kinematics of
a {\it heavy} $b$ parton inside a hadron,\refmark{\rjohn}
the event rate of single top quark production via this process
is estimated by using the method proposed in Ref.~\wuki.
The prescription proposed in Ref.~\wuki\ was shown to
be correct in leading log approximation
for the total production rate of $t$ in the $W$--gluon fusion process.
A study on how to implement a similar technique
when kinematic cuts are
applied in the event analysis is in progress.\refmark{\rjohn}
We chose the QCD scale to be $M_W$, the $W$--boson mass, and found that the
total production rate is 2 pb for $t$ and  the same rate for $\bar t$
at the Tevatron.
The parton distribution function used in our calculation is
the leading order set, Fit SL, of Morfin--Tung.\refmark\pdf
For comparison, the total production rate of $t \bar t$ via the usual QCD
process is about 16 pb.

\REF\steve{ S. Mrenna and C.--P. Yuan, Phys. Rev. {\bf D46} (1992) 1007.}
\REF\sdc{ SDC technical Design Report, preprint SDC-92-201, 1992.
It was concluded that the top quark
invariant mass has a width of 9 GeV for a 150 GeV top quark.}
\REF\branch{ After the top quark is found, one can
measure the branching ratio of $t \ra b W^+(\ra l^+\nu)$ by the
ratio of $(2l+\,jets)$ and $(1l+\,jets)$
event rates from $t \bar t$ production.
Then, a model independent measurement of the decay
width~$\width$ can be made by counting the production rate
of $t$ in the $W$--gluon fusion process. }

In Ref.~\steve, we showed that the intrinsic width of the top quark
can not be measured at the SSC and the LHC
through the usual QCD process. For instance,
the intrinsic width of a 140 GeV Standard Model top quark is
about 0.6 GeV, and the full width at half maximum of the reconstructed
top quark invariant mass is about 11 GeV after including the detector
resolution
effects by smearing the final state parton momenta. A similar conclusion was
also given from a hadron level analysis.\refmark{\sdc}
Can the top quark width~$\width$ be measured better than the factor
$11/0.6 \sim 20$ mentioned above? The answer is yes. As
pointed out in Ref.~\steve,
the width $\width$ can be measured by counting the production rate of top
quarks
from the $W$--$b$ fusion process which is {\it equivalent} to the $W$--gluon
fusion process by a proper treatment of the bottom
quark and the $W$ boson as partons inside the
hadron.\refmark{\rjohn}
 The $W$--boson which interacts with the $b$--quark to produce the top
quark can be treated as an on--shell boson
in the leading log approximation.\refmark{\effw}
 The moral is that even
under the approximations considered,
a factor of 2 uncertainty in the production
rate for this process gives a
factor of 2 uncertainty in the measurement of $\width$.\refmark{\branch}
This is still much better than what can be done
at the SSC and the LHC through the usual QCD process.
Therefore, this is an extremely important measurement to be made at
the Tevatron because it directly tests the coupling of~\tbW at the tree level.

The Effective--$W$ Approximation  has been the essential tool
used in studying the strongly interacting longitudinal $W$ system to probe the
symmetry breaking sector at the SSC.
One can learn about the validity of the Effective--$W$ Approximation prior
to supercolliders
by studying the $W$--gluon fusion process at the Tevatron.

\REF\chiral{S. Weinberg, Phys. Rev. {\bf 166} (1968) 1568;
S. Coleman, J. Wess and B. Zumino, Phys. Rev. {\bf 177} (1969) 2239;
C. Callan, S. Coleman, J. Wess and B. Zumino, Phys. Rev. {\bf 177} (1969) 2247;
M. Chanowitz, H. Georgi and M. Golden, Phys. Rev. Lett. {\bf 57} (1986) 2344.}

For heavy top quark production, the dominant subprocess is
$W_L b \ra t$, where $W_L$ is a longitudinally polarized $W$--boson.
Based on the argument of the electroweak chiral lagrangian,\refmark{\chiral}
it is likely that new physics
will show up in the interaction of a longitudinal $W$, which is equivalent to
a Goldstone boson characterizing the symmetry breaking mechanism, and a heavy
fermion. Hence, measuring the form factor of~\tbW in this process provides a
probe on the symmetry breaking sector.

As discussed above, the top quark produced from this process is about one
hundred
percent polarized due to the left--handed charged current in the SM.
This allows a test of CP violation asymmetry.\refmark{\toppol}
A detailed analysis using this process testing the CP
violation asymmetry in the decay of $t$  was
performed in Ref.~\gunion.
Furthermore, if the CP violating effects in the couplings of ~\tbW
are large, the production
rates of $ \bar p p \ra t X$ and $\bar p p\ra \bar t X$
from the $W$--gluon fusion process will be significantly different.

\FIG\fone{ $m_t$ distribution for a 140 GeV top quark at the Tevatron. It
includes both the signal and the background events
with $W^\pm \ra e^\pm \, {\rm or} \, \mu^\pm$ after the cuts~\eone\ and~\etwo.}
\FIG\ftwo{ Same as Fig.~\fone\ but with detector resolution effects
as described in~\efive.}

\REF\monte{ This correlation, arising from the polarization
of the top quark, is absent in the
full event generators such as PYTHIA or HERWIG. The
efficiency of applying this angular correlation to suppress
the background, as done in ~\etwo, is higher if
the minimum transverse momentum of the $b$ jet
as described in~\eone\ is lower. Therefore, combining the
cuts~\eone\ and~\etwo\ would guarantee the improvement of the
signal--to--background ratio.}

To show that a 140 GeV top quark produced from this process can be detected at
the Tevatron, we performed a Monte Carlo study on the $W+\,2\,jets$ mode
of the signal,
$$
 q' b \ra q t (\ra b W^+ (\ra l^+ \nu) )
\eqn\ethree
$$
with $l^+=e^+ \,{\rm or}\,\mu^+$.
After the kinematic cuts
$$
\eqalign{
 P_T^q > & 15 \, {\rm GeV}, \quad  |\eta^q| < 3.5, \cr
 P_T^l > & 15 \, {\rm GeV}, \quad  |\eta^l| < 2, \cr
 P_T^b > & 40 \, {\rm GeV}, \quad  |\eta^b| < 2, \cr
 \mynot{E_T} > & 15 \, {\rm GeV}, \quad \Delta R_{qb} > 0.7, \cr}
\eqn\eone
$$
the signal rate is about 0.11 pb.
It is important to note that the typical rapidity of
the spectator jet ($q$) in the
signal event is about 1.6 at the Tevatron.\refmark{\wgone}
The distribution of $\eta^q$ is asymmetric because the Tevatron is a
$\bar p p$ collider.
 A cut on $|\eta^q| < 3.5$ keeps almost all the signal events.
 An asymmetric cut on $P_T$ was used to suppress
the major EW-QCD background process $W+\,2\,jets$.

\REF\rnewtwo{Applying the cuts of Ref.~\parke,
we reproduced the result that for a
140 GeV top quark the $t \bar t$ background rate is about the same as the
signal rate.}

The $t \bar t$ background is not important after
vetoing the events with $\ge 3$ jets.\refmark{\parke}
We require one charged lepton and two jets,
one of which is a $b$ or $\bar b$, to pass
the kinematic cuts~\eone.  The semileptonic decay mode of the
$t \bar t$ background process
$t \bar t \ra b \bar b l^+ {l'}^- \nu {\bar\nu}'$ then gives a rate of
$\sim 0.014$ pb, where $l^+$ includes both $e^+$ and $\mu^+$, and
${l'}^-$ includes $e^-$, $\mu^-$ and $\tau^-$.
This is about a factor of 8 smaller than the signal rate.\refmark{\rnewtwo}
The event rate after the
cuts~\eone\ for one of the top quarks ($t$ or $\bar t$) decaying
semileptonically and the other hadronically is smaller than the
purely semileptonic decay of $t \bar t$ by a factor of 20.

Following the method proposed in Ref.~\wgone, the longitudinal momentum
 $P_z^\nu$
of the reconstructed (fake) neutrino in
$t \bar t$ events can be determined by requiring
the invariant mass of
the charged lepton and the reconstructed (fake) neutrino
 to be the mass of the $W$--boson.  If
a physical solution cannot be found, this event is discarded.
As expected, because the missing transverse momentum of the $t \bar t$
event is in general not that of a real neutrino from $W$ decay, neither
the transverse mass $M_T^{e \nu b}$ nor the top invariant mass $m_t$
distribution look like those of the signal event even if
a physical solution exists for $P_z^\nu$.

Hence, the dominant background is
$$
u \bar d, \, c \bar s \ra b \bar b W^+(\ra l^+ \nu).
\eqn\efour
$$
After the kinematic cuts~\eone,
this background rate is about 1.3 times the signal rate.
 The other backgrounds such as $c g \ra b g W^+$ are suppressed due
to the small CKM matrix element $|V_{ts}| \sim 0.03\, {\rm to} \,0.05$.

Because the top quark produced from the signal process
is left--handed polarized, $l^+$ tends to move against
 the moving direction of the top quark in the center--of--mass frame of $q$ and
$t$.\refmark{\toppol,\monte}
However, in the background event, the distribution of $\cos \theta_{lq}$
is almost flat after the cuts~\eone. ($\theta_{lq}=\pi-\theta_l$, where
$\theta_l$ is the the polar angle of $l^+$
in the rest frame of $t$ defined in the $q$ $t$ center--of--mass frame.)
We can further improve the signal--to--background ratio by imposing
$$
\cos \theta_{lq} > -0.4.
\eqn\etwo
$$
This was performed by using the constructed $P_z^\nu$ information to
boost into the center--of--mass of the system, we then boost
into the rest frame of the top.
After cut (2.4),
the signal--to--background ratio is
about one. There are $\sim 10$ signal events~\ethree\ with a
100 ${\rm pb}^{-1}$ luminosity and $\sqrt{S}=1.8$ TeV.
The signal event rate was reduced by a factor of 3 after applying
the cuts~\eone\ and~\etwo\ on the process~\ethree.
However, we anticipate the signal event rates will be somewhat larger than
the ones reported here after properly treating the $b$ quark as a parton
 inside the proton.\Ref\rates{ The cross section of~\ethree\
is about 1.5 pb, but the {\it true} cross
section is about 2 pb after applying the prescription given in Ref.~\wuki.}
In Fig.~\fone, we show the reconstructed
invariant mass ($m_t$) of the top quark.
The method of reconstructing $m_t$ was already presented in Ref.~\wgone.
We therefore conclude that the top quark can be detected
and studied via this process at the Tevatron.

To incorporate the effects of detector efficiencies, we smear the final state
parton momentum using a Gaussian distribution with
$$
(\Delta E / E)_l=15\% / \sqrt{E}, \quad {\rm and} \quad
(\Delta E / E)_{q,b}=50\% / \sqrt{E}.
\eqn\efive
$$
The $m_t$ distribution becomes slightly broader as shown in Fig.~\ftwo.
However, both the signal and the background rates are almost the same as those
in Fig.~\fone\ obtained with a perfect detector.

\REF\nets{ B. Denby, preprint Fermilab-conf-92-121-E, 1992.}
\REF\tten{ Based on the study done in ``Physics at Fermilab in the 1990's'',
August 15--24, 1989, Breckenridge, Colorado,
the expected $t \bar t \ra e \nu + X$ signal rate in top search
is only about 20 events with a
 $100 \, {\rm pb}^{-1}$ integrated luminosity and  $\sqrt{S}=1.8$ TeV
at the Tevatron.}

In our future publications, we will show various distributions of the signal
and
background events to illustrate that the
signal and the background events can
be further distinguished. However, we do
not suggest imposing additional cuts. Instead, we
propose other methods such as using neural networks\refmark\nets\
to obtain an even cleaner signal.
After including both charged states,
we conclude that about 20
signal events produced from the $W$--gluon
fusion process can be detected and studied to test the Standard Model and probe
new physics at the Tevatron.\refmark\tten\
In reality, the efficiency
of $b$--tagging is not perfect.
A 50\% efficiency of $b$--tagging gives a total of $\sim 10$ signal events with
 $100 \, {\rm pb}^{-1}$ integrated luminosity and  $\sqrt{S}=1.8$ TeV.
On the other hand, the Tevatron might upgrade
to $\sqrt{S}=3.6$ TeV with higher luminosity $\sim 2 \, {\rm fb}^{-1}$.
This amounts to a factor of 150 increase in the signal rate.
Hence, both D0 and CDF should make efforts to study the top quark produced
via the $W$--gluon fusion process.

\ack

C.P.Y. would like to thank
J. Collins, G. Kane, G. Ladinsky, S. Mrenna and Y.--P. Yao
for collaboration.
To them and S. Errede, J. Gunion, S. Parke, M. E. Peskin and M. Veltman,
we are grateful for discussions.
This work was funded in part by the TNRLC grant \#RGFY9240.

\refout
\figout

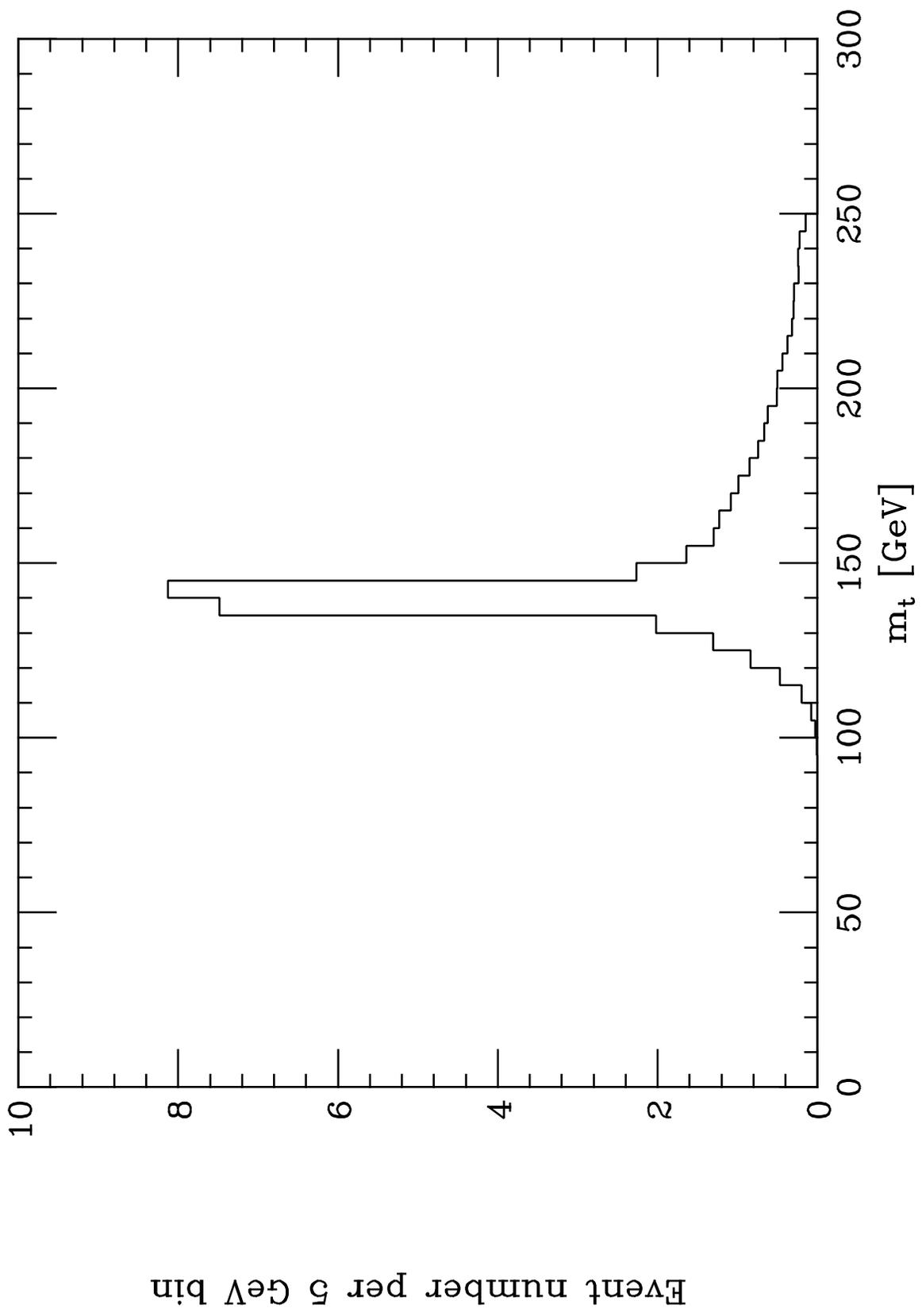
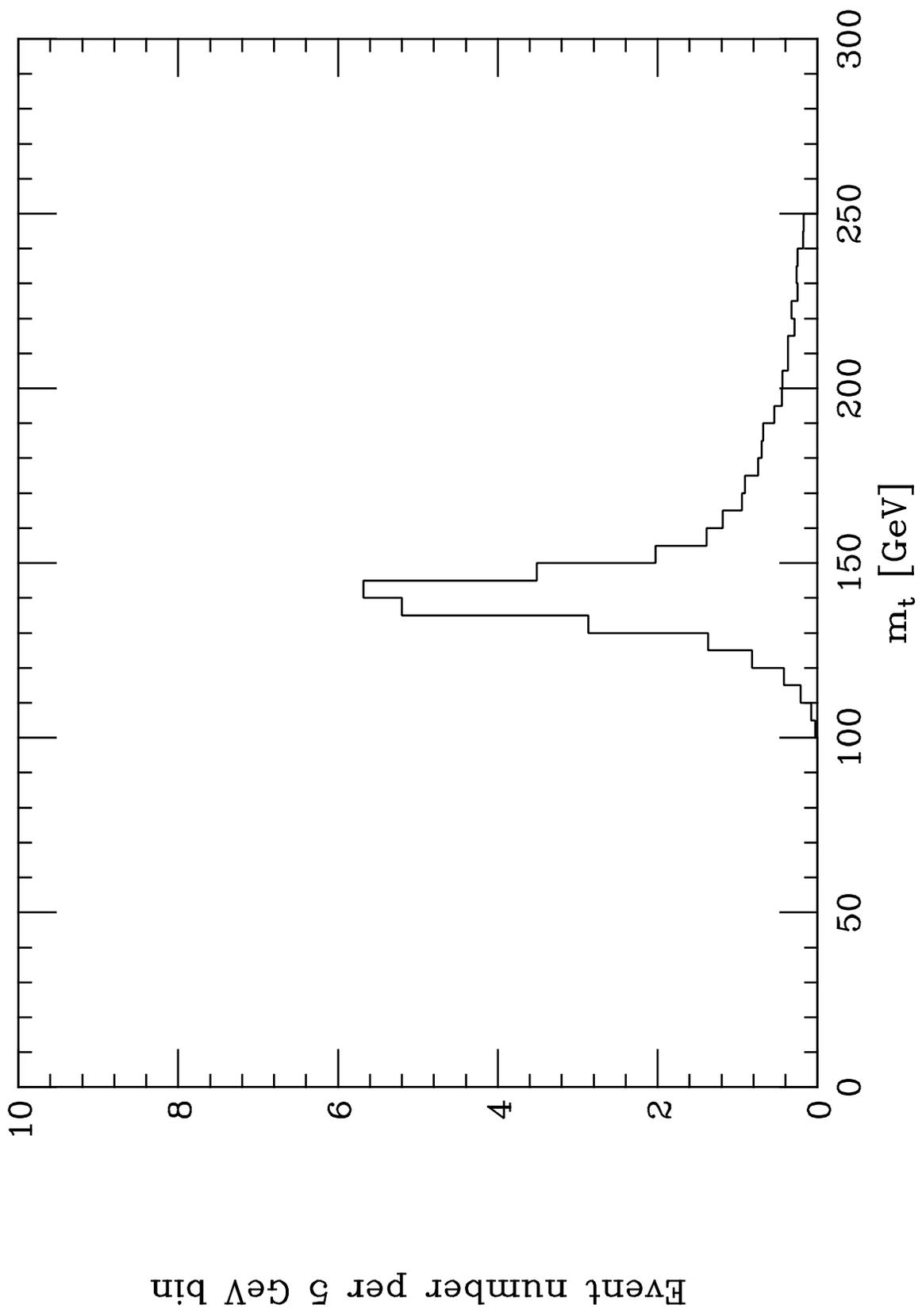
\end